\documentclass[aps,prd,twocolumn,showpacs,eqsecnum,amsmath,nofootinbib]{revtex4}
\usepackage[T1]{fontenc}
\usepackage[utf8]{inputenc}
\usepackage{lmodern}
\usepackage[english]{babel}

\usepackage{graphicx,amssymb,amsthm,amsmath,eufrak}
\usepackage{hyperref}

\def \d {{\rm d}}

\def \boldk {\mbox{\boldmath$k$}}

\begin{document}

\title{Robinson--Trautman solution with scalar hair}

\author{T. Tahamtan}
\email{tahamtan@utf.mff.cuni.cz}

\author{O. Sv\'{\i}tek}
\email{ota@matfyz.cz}
\affiliation{Institute of Theoretical Physics, Faculty of Mathematics and Physics, Charles University in Prague, V~Hole\v{s}ovi\v{c}k\'ach 2, 180~00 Prague 8, Czech Republic}

\begin{abstract}
Explicit Robinson--Trautman solution with minimally coupled free scalar field is derived and analyzed. It is shown that this solution contains curvature singularity which is initially naked but later the horizon envelopes it. We use quasilocal horizon definition and prove its existence in later retarded times using sub- and supersolution method combined with growth estimates. We show that the solution is generally of algebraic type II but reduces to type D in spherical symmetry.
\end{abstract}

\date{\today}

\maketitle

\section{Introduction}
Solutions to Einstein equations with scalar field source provide very useful tool for understanding relativity due to the simplicity of the source. Recently, it becomes progressively plausible that such fields might really exist (LHC) and potentially play a fundamental role in physics. In classical General Relativity they were used to study counterexamples to black hole no-hair theorems and in many other areas. These results were mostly based on highly symmetric solutions and it is therefore important to provide solutions with less or no symmetries to subsequently analyze if those results hold in more generic situations and are not tied to a specific symmetry.

Robinson--Trautman spacetimes represent an important class of expanding nontwisting and nonshearing solutions \cite{RobinsonTrautman:1960,RobinsonTrautman:1962,Stephanietal:book} describing non-spherical generalizations of black holes. In general, they do not posses any Killing vectors thus providing important solutions devoid of symmetry. Many properties of this family in four dimensions have been studied, especially in the last 25 years. In particular, the existence, asymptotic behaviour and global structure of vacuum Robinson--Trautman spacetimes of type~II with spherical topology were investigated by Chru\'{s}ciel and Singleton \cite{Chru1,Chru2,ChruSin}. Robinson--Trautman solutions were shown to exist for generic, arbitrarily strong smooth initial data for all positive retarded times, and to converge asymptotically to corresponding Schwarzschild metric. Extensions across the ``Schwarzschild-like'' future event horizon can only be made with a finite order of smoothness. These results were generalized in \cite{podbic95,podbic97} to Robinson--Trautman vacuum spacetimes with cosmological constant. These cosmological solutions settle down to a Schwarzschild--(anti-)de Sitter solution at large times $u$. Finally, the  Chru\'{s}ciel--Singleton analysis was extended to Robinson-Trautman spacetimes including matter, namely pure radiation \cite{BicakPerjes,PodSvi:2005}, showing that they approach the spherically symmetric \mbox{Vaidya--(anti-)de~Sitter} metric. Generally, the solutions of this family settle down to physically important solutions. The location of the horizon together with its general existence and uniqueness for the vacuum Robinson--Trautman solutions has been studied by Tod \cite{tod}. Later, Chow and Lun \cite{chow-lun} analyzed some other useful properties of this horizon and made numerical study of both the horizon equation and Robinson--Trautman equation. These results were later extended to nonvanishing cosmological constant \cite{PodSvi:2009}. The anisotropy of Robinson-Trautman horizon and its associated asymptotic momentum was also used in the analytic explanation of an "antikick" appearing in numerical studies of binary black hole mergers \cite{rezzolla}.

Robinson--Trautman spacetimes (containing aligned pure radiation and a cosmological constant) were also generalized to any dimension \cite{podolsky-ortaggio}. Existence of horizons was subsequently analyzed in \cite{Svitek2011}. Finally, Robinson--Trautman solutions with p-form fields in arbitrary dimension were derived recently \cite{ortaggio}. One of the results mentioned therein rules out the existence of aligned scalar field (where alignment refers to the gradient of the field) for generic Robinson--Trautman case.

The solutions for "stringy" Robinson--Trautman spacetime corresponding to Einstein--Maxwell--dilaton system were obtained in \cite{Guven}. Recently, scalar field solutions for Einstein--Maxwell--Lambda system with a conformally coupled scalar field belonging to Pleba\'{n}ski--Demia\'{n}ski family (containing type D solutions of Robinson--Trautman class) were derived in \cite{Maeda}. 

\section{Vacuum Robinson--Trautman metric and field equations}
\label{RTmetricsec}

The general form of a vacuum Robinson--Trautman spacetime can be given by the following line element \cite{RobinsonTrautman:1960,RobinsonTrautman:1962,Stephanietal:book,GriffithsPodolsky:book}
\begin{equation}\label{RTmetric}
\d s^2 = -2H\,\d u^2-\,2\,\d u\,\d r + \frac{r^2}{\tilde{P}^2}\,(\d y^{2} + \d x^{2}),
\end{equation}
where ${2H = \Delta(\,\ln \tilde{P}) -2r(\,\ln \tilde{P})_{,u} -{2m/r} -(\Lambda/3) r^2}$,
\begin{equation}\label{Laplace}
\Delta\equiv \tilde{P}^2(\partial_{xx}+\partial_{yy}),
\end{equation}
and $\Lambda$ is the cosmological constant. The metric depends on two functions, ${\,\tilde{P}(u,x,y)\,}$ and ${\,m(u)\,}$, which satisfy the nonlinear Robinson--Trautman equation
\begin{equation}
\Delta\Delta(\,\ln \tilde{P})+12\,m(\,\ln \tilde{P})_{,u}-4\,m_{,u}=0\,.
\label{RTequationgen}
\end{equation}
The function $m(u)$ might be set to a constant by suitable coordinate transformation for vacuum solution.

The spacetime admits a geodesic, shearfree, twistfree and expanding null congruence generated by ${\boldk=\partial_r}$. The coordinate $r$ is an affine parameter along this congruence, $u$~is a retarded time coordinate, and $x,y$ are spatial coordinates spanning transversal 2-space with their Gaussian curvature (for ${r=1}$) being given by
\begin{equation}
\mathcal{K}(x,y,u)\equiv\Delta(\,\ln \tilde{P})\,.
\label{RTGausscurvature}
\end{equation}
For general fixed values of $r$ and $u$, the Gaussian curvature is ${\mathcal{K}/r^2}$ so that, as ${r\to\infty}$, they become locally flat.

\section{Solution coupled to a scalar field}
We consider the following action, describing a scalar field minimally coupled to gravity, 
\begin{equation}\label{action}
S=\int d^{4}x \sqrt{-g}[\mathcal{R}+\nabla_{\mu}\varphi \nabla^{\mu} \varphi]
\end{equation}
where $\mathcal{R}$ is the Ricci scalar for the metric $g_{\mu \nu}$. The massless scalar field $ \varphi $ is supposed to be real and we use units in which $c=\hbar=8 \pi G=1$. By applying the variation with respect to the metric for the action (\ref{action}), we get Einstein equations
\begin{equation}\label{field equations}
\mathcal{R}_{\mu \nu}-\frac{1}{2}g_{\mu \nu}\mathcal{R}=T_{\mu \nu}.
\end{equation}
The energy momentum tensor generated by the scalar field is given by
\begin{equation} \label{energy-momentum}
T_{\mu \nu}=\nabla_{\mu}\varphi \nabla_{\nu}\varphi-\frac{1}{2}g_{\mu \nu}g^{\alpha\beta}\nabla_{\alpha}\varphi \nabla_{\beta}\varphi
\end{equation}
and the scalar field must satisfy corresponding field equation
\begin{equation}\label{box}
\Box \varphi(u,r)=0
\end{equation}
where $\Box$ is a standard d'Alembert operator for our metric (\ref{ourmetric}).

For the matter of convenience we will be looking for the metric in the following form
\begin{eqnarray}\label{ourmetric}
 \d s^2&=&-2(H(u,r)+K(u,x,y))\,\d u^2-2\,\d u\d r \nonumber\\
&& +\frac{R(u,r)^2}{P(x,y)^2}(\d x^2+\d y^2)
 \end{eqnarray}
 The scalar field is assumed to be function of $u$ and $r$ only ($\varphi(u,r)$). The dependence on $r$ means that the scalar field is not aligned and thus is not ruled out by the results of \cite{ortaggio}.  The nontrivial components of the Ricci tensor corresponding to the metric (\ref{ourmetric}) are
\begin{eqnarray}
\mathcal{R}_{uu}&=&2\left(2\frac{R_{,r}}{R}H_{,r}+H_{,rr}\right)(H+K)+2\frac{R_{,r}}{R}(H+K)_{,u} \nonumber\\
&&-\frac{2}{R}(R_{,u}H_{,r}+R_{,uu})+\frac{P^2}{R^2}(K_{,xx}+K_{,yy})\nonumber\\
\mathcal{R}_{rr}&=&-2\frac{R_{,rr}}{R} \\
\mathcal{R}_{ru}&=&\mathcal{R}_{ur}=2\frac{R_{,r}H_{,r}-R_{,ru}}{R}+H_{,rr}\nonumber\\
\mathcal{R}_{xx}&=&\mathcal{R}_{yy}=-\frac{1}{P^2}\left\{k(x,y)+2(H+K)(RR_{,r})_{,r}+\right. \nonumber\\
&&\left.+2RR_{,r}H_{,r}-2(RR_{,u})_{,r}\right\}\nonumber
\end{eqnarray}
where as usual $()_{,x^i}=\frac{\partial }{\partial x^i}()$ and
\begin{equation}
k(x,y)=\Delta(\,\ln P(x,y))
\end{equation}
where $\Delta$ is still given by expression (\ref{Laplace}) with $\tilde{P}$ replaced by $P$.

We will use the following form of equations equivalent to Einstein equations (\ref{field equations}) coupled to energy momentum tensor (\ref{energy-momentum})
\begin{equation}
\mathcal{R}_{\mu \nu}=\varphi_{,\mu}\varphi_{,\nu}=\left(
\begin{array}{cccc}
\varphi_{,u}^2 & \varphi_{,u}\varphi_{,r} &0&0\\ 
\varphi_{,u}\varphi_{,r} & \varphi_{,r}^2&0&0\\
0&0&0&0\\
0&0&0&0
\end{array}
\right)
\end{equation}
From the above equations describing gravitational field and field equation for the scalar field (\ref{box}) we obtain the following expressions for unknown metric functions and scalar field 
\begin{eqnarray}\label{solution}
H(u,r)&=&\frac{r}{2U(u)}\frac{\partial U(u)}{\partial u} \nonumber\\
R(u,r)&=&\sqrt{\frac{U(u)^{2}r^2-C_{0}^2}{U(u)}}  \nonumber\\
K(u,x,y)&=&\frac{k(x,y)}{2U(u)} \\
\varphi(u,r)&=&\frac{1}{\sqrt{2}}\ln{\left\{ \frac{U(u)r-C_{0}}{U(u)r+C_{0}}\right \}}\nonumber\\
\Delta k(x,y)&=&{\alpha^2} \nonumber\\
U(u)&=&\gamma e^{\omega^2 u^2+\eta u},\nonumber 
\end{eqnarray}
in which $ C_{0} \neq 0, \alpha, \eta, \gamma, \omega $ are constants and $\omega=\frac{\alpha}{2C_{0}}$. In the following we will assume $ C_{0} > 0, \alpha > 0, \eta > 0, \gamma > 0$ for simplicity of discussion.

\section{Properties of the solution}
First, we should ensure that our solution really belongs to the Robinson-Trautman family. This is simply confirmed by studying the properties of a null congruence generated by vector $\mathbf{l}=\partial_{r}$. Such congruence is geodesic, nontwisting, nonshearing and its expansion is given by
\begin{equation}\label{lexpansion}
	\Theta_{\mathbf{l}}=2\frac{R_{,r}}{R}=\frac{2\,U(u)^{2}r}{U(u)^{2}r^{2}-C_{0}^{2}}.
\end{equation}
Evidently, the above expression is positive only for $r > \frac{C_{0}}{U(u)}$ which may seem not satisfactory. However by inspecting the Kretschmann scalar 
\begin{equation}
	\kappa \sim \frac{1}{R(u,r)^{8}}
\end{equation}
and using (\ref{solution}) we immediately see that the geometry has singularities for $r=\pm r_{0}=\pm\frac{C_{0}}{U(u)}$. Naturally, we are led to constrain the range of coordinates to $r\in\left(\frac{C_{0}}{U(u)},\infty\right)$. In this range the expansion (\ref{lexpansion}) is everywhere positive, diverges at the singularity and approaches zero at infinity (as $r\to\infty$). Also, one can check from the line element that the singularity is a standard pointlike one. Due to the asymptotic behaviour of function $U(u)$ (see (\ref{solution})) the singularity tends to $r_{0}=0$ as $u\to\infty$. The singularity appears due to the divergence of the scalar field and its energy momentum tensor.

Asymptotically ($u\to\infty$), the scalar field itself is vanishing everywhere outside the singularity (see (\ref{solution})) while it diverges at $r=0$. So there would be no scalar hair left outside when the spacetime settles down to the final state. Indeed, our geometry approaches the original Robinson--Trautman form (\ref{RTmetric}) for $u\to\infty$ when we define $\tilde{P}(u,x,y)=P(x,y)/U(u)$. In this case one can apply the Chru\'{s}ciel--Singleton analysis \cite{Chru1,Chru2,ChruSin} of asymptotic behaviour to recover the spherical symmetry of the final state which neccesarrily points to Schwarzschild solution.

When the singularity is present in our solution we will investigate if it is covered by a horizon. Due to dynamical nature of the spacetime it is preferable to use the quasilocal definitions of horizon --- apparent horizon \cite{hawking-ellis}, trapping horizon \cite{hayward} or dynamical horizon \cite{krishnan}. The basic {\it local} condition is shared by all the standard horizon definitions: these horizons are sliced by marginally trapped surfaces with vanishing expansion of outgoing (ingoing) null congruence orthogonal to the surface. We will be looking for the horizon hypersurface in the following form
\begin{equation}
	r=\mathcal{M}(u,x,y)
\end{equation}
and study the expansion of compact slices of such hypersurface given by $u=u_{0}=const.$ (with $\mathcal{M}(u_{0},x,y)=M(x,y)$). The requirement of compactness necessarily means that the two-spaces spanned by $x$ and $y$ are compact as well. We construct null vector fields orthogonal to surface $r=M(x,y)$
\begin{eqnarray}\label{null-vectors}
	\mathbf{l}&=&\partial_{r}\\
	\mathbf{k}&=&\partial_{u}+\left[\frac{P^{2}}{2R^{2}}(M_{x}^{2}+M_{y}^{2})-(H+K)\right]\partial_{r}+\nonumber\\
	&&+\frac{P^{2}}{R^{2}}(M_{x}\partial_{x}+M_{y}\partial_{y})
\end{eqnarray}
that satisfy normalization condition $\mathbf{l}\cdot \mathbf{k}=-1$. From the geometry of the situation one can deduce that congruence $\mathbf{l}$ is outgoing while $\mathbf{k}$ is ingoing. The expansion of the congruence generated by $\mathbf{l}$ is always positive (see (\ref{lexpansion}) and  the discussion beneath) so we are looking for vanishing of expansion related to the other congruence $\mathbf{k}$. These two conditions ($\Theta_{\mathbf{l}}>0$ and $\Theta_{\mathbf{l}}=0$) mean that we are looking for the past horizon according to the definition given by \cite{hayward}. The second expansion is given by
\begin{eqnarray}\label{kexpansion}
	\Theta_{\mathbf{k}}&=&\frac{1}{R^{2}}\left[\Delta M-(\ln{R})_{,r}(\nabla M \cdot \nabla M)-\right.\nonumber\\
	&&\left.-(K+H)(R^{2})_{,r}+(R^{2})_{,u}\right],
\end{eqnarray}
where Laplace operator and scalar product denoted by dot correspond to metric $h_{ij}dx^{i}dx^{j}=\frac{1}{P(x,y)^{2}}(dx^{2}+dy^{2})$ on the space $\Sigma$ spanned by $x,y$. So the horizon is given by the solution of the following quasilinear elliptic partial differential equation
\begin{eqnarray}\label{horizon}
	\left\{\Delta M-(\ln{R})_{,r}(\nabla M \cdot \nabla M)-(K+H)(R^{2})_{,r}\right.&&\nonumber\\
	\left.+(R^{2})_{,u}\right\}\vert_{r=M(x,y)\& u=u_{0}}=0&&
\end{eqnarray}
where all dependence on $r$ is replaced by the function $M(x,y)$ and $u$ is evaluated to arbitrary constant value $u_{0}$. 

It is impossible to solve this equation generally but fortunately we can get some useful information about the existence of solution using the technique developed for the case of Robinson--Trautman spacetime in higher dimensions \cite{Svitek2011}. The proof of existence of the solution to the same type of quasilinear equation ($\Delta u=F(x,u,\nabla u)$) was given there by combining several steps motivated by \cite{Kuo} and using results from \cite{Besse,Boccardo,Gilbarg-Trudinger}. The main issues were to provide an estimate for the function $F$ of the form $\vert F\vert \leq B(u)(1+\vert{\nabla u}\vert^2)$ (where $B(u)$ is increasing function on ${\mathbb R}^{+}$), and to show the existence of a sub- and a super-solution \cite{footnote} $u^{-}\leq u^{+}$, $u^{\pm}\in C^{1,\beta}(\Sigma)\cap L^{\infty}(\Sigma)$ (here $C^{1,\beta}(\Sigma)$ are H\"{o}lder continuous functions of some suitable index $\beta$). Then we know there is a solution $u \in C^{2,\iota}(\Sigma)$ (for some $\iota$) satisfying $u^{-}\leq u \leq u^{+}$.

In our case, to provide an estimate of the form $\vert M\vert \leq B(M)(1+\vert{\nabla M}\vert^2)$ (the norm is taken with respect to the two-dimensional metric $h_{ij}$) for the horizon equation (\ref{horizon}) when considered in the form $\Delta M=F(x,y,M,\nabla M)$ where
\begin{equation}\label{Function}
	F=(\ln{R(u_{0},M)})_{,r}\vert\nabla M\vert^{2}+k(x,y)M-\frac{C_{0}^{2}U_{,u}(u_{0})}{U^{2}(u_{0})}
\end{equation}
one has to deal with the singular behaviour of $(\ln{R})_{,r}$ at $r=\frac{C_{0}}{U(u)}$. We can do this either by removing the vicinity of singularity from our domain $r\in \mathbf{R}^{+}\setminus \left(\frac{C_{0}}{U(u)}(1-\delta),\frac{C_{0}}{U(u)}(1+\delta)\right)$ or by continuing (with some appropriate smoothing) the divergent function on the problematic interval $\left(\frac{C_{0}}{U(u)}(1-\delta),\frac{C_{0}}{U(u)}(1+\delta)\right)$ with a constant value it attains on the boundary of the interval. Now, with all the coefficients of the equation finite one can construct the bounding function $B(u)$ easily and thus we can proceed to the construction of sub- and supersolutions $M^{\pm}$.

First, we note that due to the selection of sign for the free constants made at the end of previous section ($ C_{0} > 0, \alpha > 0, \eta > 0, \gamma > 0$) we obtain $U_{,u}>0$ if we restrict our attention to retarded time region $u\in (-\frac{\eta}{2\omega^{2}}, \infty)$. We can then understand our solution as being given by initial conditions specified at $u_{in}=-\frac{\eta}{2\omega^{2}}$ which corresponds to usual understanding of Robinson--Trautman solution. As usual, we are looking for constant sub- and supersolutions but we are unable to provide them independently of the value of $u_{0}$. Generally, we can find the sub- and supersolutions in the following cases:
\begin{itemize}
	\item $u_{0} < u_{1}=\frac{\min({k(x,y)})-\max({k(x,y)})\delta-C_{0}\eta}{2C_{0}\omega^{2}}$
		\begin{eqnarray}
			M^{-}&=&0\\
			M^{+}&=&\frac{C_{0}}{U(u_{0})}(1-\delta)
		\end{eqnarray}
	\item $u_{0} > u_{2}=\frac{\max({k(x,y)})(1+\delta)-C_{0}\eta}{2C_{0}\omega^{2}}$
		\begin{eqnarray}
			M^{-}&=&\frac{C_{0}}{U(u_{0})}(1+\delta)\\
			M^{+}&=&\frac{C_{0}^{2}U_{,u}(u_{0})}{\min({k(x,y)})U^{2}(u_{0})}
		\end{eqnarray}
\end{itemize}
Both bounds $u_{1}$ and $u_{2}$ are in the restricted range of coordinate $u$. Evidently, the first case would provide existence of solution only beneath the position of singularity (or, in other words, inside the singularity) which is irrelevant and moreover we have already restricted the range of $r\in\left(\frac{C_{0}}{U(u)},\infty\right)$. In the second case, one can easily check that the necessary condition $M^{-}\leq M^{+}$ is indeed satisfied for $u_{0} > u_{2}$ and we certainly have a horizon given by $r=M(x,y)$ where $M^{-}\leq M(x,y) \leq M^{+}$. Note that we suppose that $\min({k(x,y)})>0$ for the last estimate $M^{+}$ to be valid.

If we allow $k_{min}\equiv\min({k(x,y)})\leq 0$ (we define accordingly $k_{max}\equiv\max({k(x,y)})$) we are unable to provide constant supersolution in the case $u_{0} > u_{2}$. Instead we can use the knowledge of how Laplace operator acts on $k(x,y)$ (\ref{solution}) and the observation that the first term of the definition of function $F$ (\ref{Function}) is always positive to provide the following non-constant supersolution
\begin{equation}
	M^{+}=c[k_{max}-k(x,y)]+\frac{C_{0}}{U(u_{0})}(1+\delta)
\end{equation}
where
\begin{equation}
	c=\frac{C_{0}\left[C_{0}U_{,u}(u_{0})-k_{min}U(u_{0})(1+\delta)\right]}{U^{2}(u_{0})(\alpha^{2}+k_{max}k_{min}-k_{min}^{2})}.
\end{equation}
This estimate works if $(\alpha^{2}+k_{max}k_{min}-k_{min}^{2})>0$.

Even-though our solution possesses singularity at any retarded time $u$ this singularity appears to be initially naked and the horizon develops only in later time.

\section{Algebraic type of the solution}
Now, we would like to see if the geometry of our spacetime is sufficiently general. Since vacuum Robinson--Trautman spacetime is generally of algebraic type II we would like our solution to be at least of the same type and not more special. Our preferred tetrad for determining the Weyl scalars of our solution is given by different null vectors compared to (\ref{null-vectors})
\begin{eqnarray}
	\mathbf{\tilde{l}}&=&\partial_{r}\nonumber\\
	\mathbf{\tilde{k}}&=&\partial_{u}-(H+K)\partial_{r}\\
	\mathbf{m}&=&\frac{P}{\sqrt{2}R}(\partial_{x}+I\partial{y})\nonumber
\end{eqnarray}
where $I$ is a complex unit. The Weyl spinor computed from this tetrad has only the following nonzero components
\begin{eqnarray}\label{Weyl}
	\Psi_{0}&=&\frac{1}{4UR^{2}}\left[\frac{1}{2}P(k_{,yy}-k_{,xx}+Ik_{,xy})-\right.\nonumber\\
	&&\left.-(k_{,x}-Ik_{,y})(P_{,x}-IP_{,y})\right]\nonumber\\
	\Psi_{1}&=&\frac{\sqrt{2}PR_{,r}}{UR^{2}}(k_{,x}-Ik_{,y})\\
	\Psi_{2}&=&\frac{1}{6UR^{2}}\left[Uk-(U_{,u}r+k)(RR_{,rr}R_{,r}{}^{2})-\right.\nonumber\\
	&&\left.-2URR_{,ru}+(RU_{,u}+2UR_{,r})R_{,r}\right]\nonumber
\end{eqnarray}
Now, we can easily determine the type irrespective of possible non-optimal choice of tetrad by using the review of explicit methods for determining the algebraic type in \cite{Zakhary} that are based on \cite{Penrose}. Namely, when we use invariants
$$I=\Psi_{0}\Psi_{4}-4\Psi_{1}\Psi_{3}+3\Psi_{2}^{2},\ J={\rm det}\left(\begin{array}{ccc}
\Psi_{4} & \Psi_{3} & \Psi_{2}\\
\Psi_{3} & \Psi_{2} & \Psi_{1}\\
\Psi_{2} & \Psi_{1} & \Psi_{0}
\end{array}\right)$$
we can immediately confirm that $I^{3}=27J^{2}$ is satisfied so that we are dealing with type II or more special. At the same time generally $IJ\neq 0$ so it cannot be just type III. Moreover, the spinor covariant $R_{ABCDEF}$ has nonzero components
\begin{eqnarray}
	R_{000000}&=&\Psi_{1}(3\Psi_{0}\Psi_{2}-2\Psi_{1}^{2})\\
	R_{000001}&=&\frac{1}{2}\Psi_{2}(3\Psi_{0}\Psi_{2}-2\Psi_{1}^{2})
\end{eqnarray}
which means that generally the spacetime cannot be of type D. So indeed our scalar field solution is of the most general type possible for the Robinson-Trautman vacuum class. Which does not mean that there cannot be a scalar field solution of type I. Moreover, inspecting the components of the Weyl spinor (\ref{Weyl}) one concludes that in the special case of $k(x,y)=const > 0$ (constant positive Gaussian curvature of compact two-space spanned by $x,y$) the algebraic type becomes D consistent with spherical symmetry. Finally, since $\Psi_{1}=0$ implies $\Psi_{0}=0$ we cannot have all components of spinor covariant $Q_{ABCD}$ (see \cite{Penrose}and \cite{Zakhary}) vanishing while having nonvanishing Weyl spinor. This means that our family of solutions does not contain type N geometries.

\section{Conclusion and final remarks}
We have derived a Robinson--Trautman spacetime with minimally coupled free scalar field. We have shown that it has a singularity for all retarded times created by the divergence of the scalar field therein. This singularity is initially (with respect to retarded time) naked and only later becomes covered by the quasilocal horizon. Note that the energy momentum tensor of the free minimally coupled scalar field trivially satisfies null energy condition (as well as weak and strong ones) and the naked singularity at the beginning of the evolution is probably caused by a slow buildup of effective energy density caused by the scalar field at the singularity position which is enough to form the singularity but not enough to envelop it in horizon initially. This behaviour suggests similarity with the appearance of a naked curvature singularity in Vaidya spacetime with linear mass function. The naked singularity appears there initially depending on the speed of growth of mass \cite{WaughLake:1986} and later becomes covered by horizon as well. From the properties of both null congruences orthogonal to the horizon we deduced that we are dealing with past horizon which is natural for standard (retarded) form of Robinson--Trautman spacetime. 

Our solution is asymptotically flat, contains a black hole (at least in the later stage of development) and has a scalar field, so one is naturally interested in its connection with the no-hair theorems (see \cite{Herdeiro} for current review). As recently shown \cite{Graham}, for stationary black hole spacetimes there are no scalar hairs (even for time-dependent scalar field) which means that the dynamical nature of Robinson--Trautman family is truly needed for our solution to be feasible. Also, we have shown that the scalar field vanishes outside the black hole in infinite retarded time limit when the geometry settles down to the final state --- Schwarzschild black hole.

Finally, we have proved that our geometry is of algebraic type II (the most general type for vacuum Robinson--Trautman spacetimes) and if we restrict to spherically symmetric case it is of type D. However, the type N subcase is not possible for our solution.

\begin{acknowledgments}
This work was supported by grant GA\v{C}R 14-37086G.
\end{acknowledgments}

\end{document}